\documentclass[a4paper]{jpconf}
\usepackage{graphicx,iopams}
\begin{document}
\title{Event-by-event study of CR composition \\ with the SPHERE experiment \\ using the 2013 data}

\author{R.A. Antonov$^{1}$, T.V. Aulova$^{1}$, E.A. Bonvech$^{1}$, D.V. Chernov$^{1}$, \\ T.A. Dzhatdoev$^{1,*}$, Mich. Finger$^{2,3}$, Mir. Finger$^{2,3}$, V.I. Galkin$^{4,1}$, \\ D.A. Podgrudkov$^{4,1}$, T.M. Roganova$^{1}$}

\address{$^{1}$ Skobeltsyn Institute of Nuclear Physics, Lomonosov Moscow State University, Leninskie gory 1-2, 119991 Moscow, Russia \\
$^{2}$ Faculty of Mathematics and Physics, Charles University, Ovocn trh 3-5, 116 36 Praha 1, Prague, Czech Republic \\
$^{3}$ Joint Institute for Nuclear Research, Joliot-Curie 6, 141980 Dubna, Moscow region, Russia \\
$^{4}$ Faculty of Physics, Lomonosov Moscow State University, Leninskie gory 1-2, 119991 Moscow, Russia}
\ead{$^{*}$timur1606@gmail.com}

\begin{abstract}
We present an event-by-event study of cosmic ray (CR) composition with the reflected Cherenkov light method. The fraction of CR light component above 5 $PeV$ was reconstructed using the 2013 run data of the SPHERE experiment which observed optical Vavilov-Cherenkov radiation of extensive air showers, reflected from snow surface of Lake Baikal. Additionally, we discuss a possibility to improve the elemental groups separability by means of multidimensional criteria.
\end{abstract}

\section{Introduction}

The study of the superhigh energy ($E > 10^{15} eV$ = 1 $PeV$) cosmic ray (CR) composition is the most difficult experimental problem of CR physics. Indeed, even for the most basic primary nuclei mass-dependent quantity, the mean logarithmic mass number $<lnA>$, the spread of results is very high (almost from proton to Iron at some energies) \cite{tsu08}. A somewhat newer review, at the cost of ignoring a part of older works, shows lesser, but still considerable, scatter of $<lnA>$ values \cite{kam12}. Available information on the elemental groups spectra \cite{ant05}--\cite{ape13b} at $E>$ 3 $PeV$ is almost exclusively due to particle detectors relying on the electron and muon numbers measurement \cite{kam12}; this method is highly model dependent \cite{ape14}. Moreover, results obtained with event-by-event techniques are very scarce \cite{ant02}, and most of the above-mentioned works \cite{ant05}--\cite{ape13b} relied on the deconvolution method, that may suffer from the ill-posedness of the inverse problem \cite{ter07}.

In the present work we describe an event-by-event study of CR composition using the 2013 run data of the SPHERE experiment. The results of this Cherenkov experiment are much less dependent on high-energy hadronic model than for the case of particle detectors and, additionally, are expected to be more stable than ones obtained with the deconvolution method. The datasample used in our analysis is considered in section 2. The method employed for the composition study is described in section 3. In section 4 the result of the analysis --- the fraction of CR light component above 5 $PeV$ --- is presented. Finally, in section 5 we show that it is possible to enhance the sensitivity of the method to the primary nuclei mass number with multidimensional criteria.

\section{The datasample and low-level data analysis}

The SPHERE experiment, stage 2 (2008-2013), employed the SPHERE-2 balloon-borne detector \cite{ant15a} to observe optical Vavilov-Cherenkov radiation ("Cherenkov light") of extensive air showers (EAS), reflected from snow surface of Lake Baikal. The foundations of the method were discussed in \cite{chu74}. Many details on the detector hardware, observation conditions, simulations, and experimental data analysis could be found in \cite{ant15a}; for brief review see \cite{ant15b}. During the 2013 run, the typical observation altitude (measured by GPS) was in range $H$= 400--700 $m$ above the snow surface. In total, 3813 events were recorded; 459 of them were recognized as EAS events (for the low-level experimental data analysis methods, see \cite{ant15a}, section 5.4). For the next step of the analysis, the lateral distribution function (LDF) reconstruction, we employed a new code, that is generally similar to the code used in our previous works \cite{ant13}, \cite{ant15a}, but was written in a completely independent manner. This circumstance allows us to test the robustness of our results against those obtained with the "old" LDF reconstruction method \cite{ant13}.

In total, 421 LDF were obtained with the new method out of the 459 events. The zenith and azimuthal angles of these showers $(\theta,\varphi)$ were measured under an assumption that the shower's front is a plane. Out of the 421 LDF 354 events have estimated zenith angles $\theta< 40^{\circ}$ and reconstructed axis distance to the center of the detector's field-of-view (FOV) projection to the observation surface $R< R_{Max}$; $R_{Max}= R_{FOV}$+30 $m$, where $R_{FOV}= 0.4903\cdot H$ is the FOV projection radius. The present analysis relies on a subsample of the late 354 LDF, 328 events. The simulation, however, was perfomed for the full sample of showers with $\theta< 40^{\circ}$ and $R< R_{Max}$, that corresponds to 354 events. Given that the experimental sample is nearly complete, in what follows we neglect all possible associated biases.

\section{The method to separate elemental groups}

\subsection{The LDF steepness parameter}

For composition study we follow the general approach developed in \cite{ano09}--\cite{ant09}, \cite{ant13}. Using a sample of model LDF, we define the criteria of the elemental groups separation for different primary energies, zenith angles and observation altitudes. As in \cite{ano09}--\cite{ant09}, \cite{ant13}, the parameter sensitive to the primary nuclei mass is the LDF "steepness", i.e. the ratio of the number of Cherenkov photons in the circle with the radius of 70 $m$ with the center in the LDF's axis to the same number in the concentric ring with the radii of 70 $m$ and 140 $m$.

\subsection{Simulations}

Simulations of EAS development were carried out for primary energies $E_{MC}$= 10, 30, 100 $PeV$ by means of full direct Monte Carlo (MC) method using the CORSIKA code \cite{hec98} with the QGSJET-I high energy hadronic model \cite{kal97} and the GHEISHA low energy hadronic model \cite{fes85}. The detector response database, that consists of a large number ($\sim 3\cdot 10^{5}$) of model showers, was calculated using the Geant4 code \cite{ago03}. Each "CORSIKA shower" was used multiple times with different axis coordinates. The axis coordinates were uniformly (randomly) distributed over big square with dimensions $1.5\cdot H\times 1.5\cdot H$.

As well, the instrumental trigger response was simulated for a range of energies $E_{Trig}=$5-200 $PeV$, and for a wide range of other conditions. For $E_{Trig}=$5--17.3 $PeV$, the 10 $PeV$ CORSIKA showers were used, and for $E_{Trig}>$17.3 $PeV$ --- the 30 PeV showers. The model showers with $E_{Trig}/E_{MC}= K$ were obtained by multipying the corresponding responses to the factor $K$. We have verified that the $E_{MC}$ values discreteness has a negligible impact on results of the trigger response simulation.

\subsection{Estimation of shower parameters}

\begin{figure}[t]
\includegraphics[width=18pc]{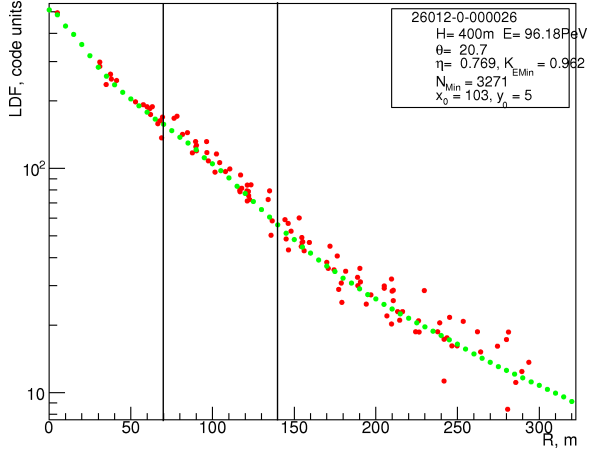}\hspace{2pc}%
\begin{minipage}[b]{18pc}\caption{\label{label} An example of the individual model LDF (red circles) together with a "composite model LDF" (green circles) that fits the individual LDF. The energy of primary nucleus was set to 100 $PeV$. The legend shows the values of some "true MC" parameters; $(x_0,y_0)$ $[m]$ is the axis position coordinates with respect to the center of the detector's FOV projection to the observation surface. The zenith angle value $\theta$ is measured in $^{\circ}$. As well, the reconstructed energy value (96.18 $PeV$) is shown.}
\end{minipage}
\end{figure}

\begin{figure}[t]
\begin{minipage}{18pc}
\includegraphics[width=18pc]{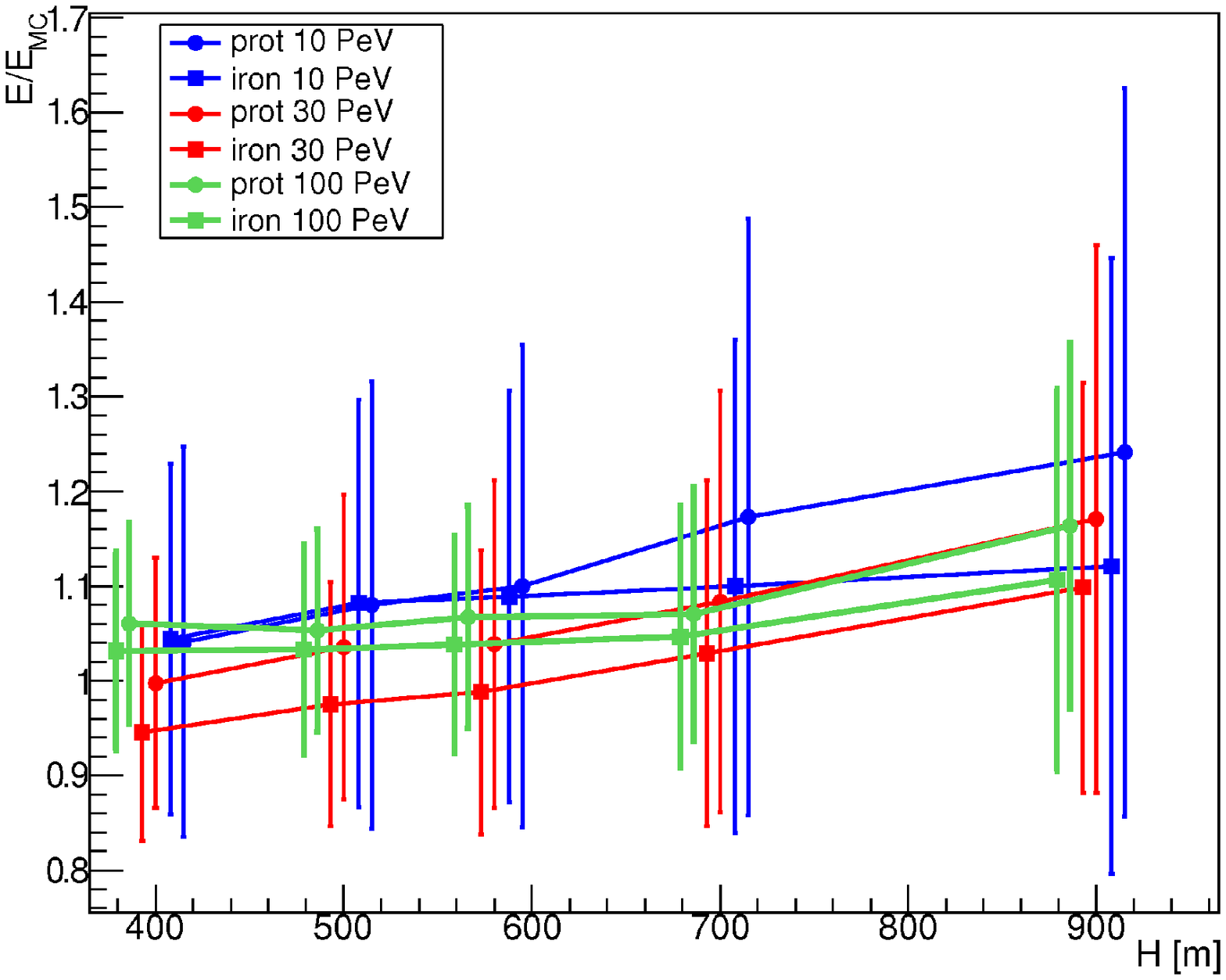}
\caption{\label{label} Reconstructed energy for model showers with various primary nuclei, primary energies and observation altitudes.}
\end{minipage}\hspace{2pc}%
\begin{minipage}{18pc}
\includegraphics[width=18pc]{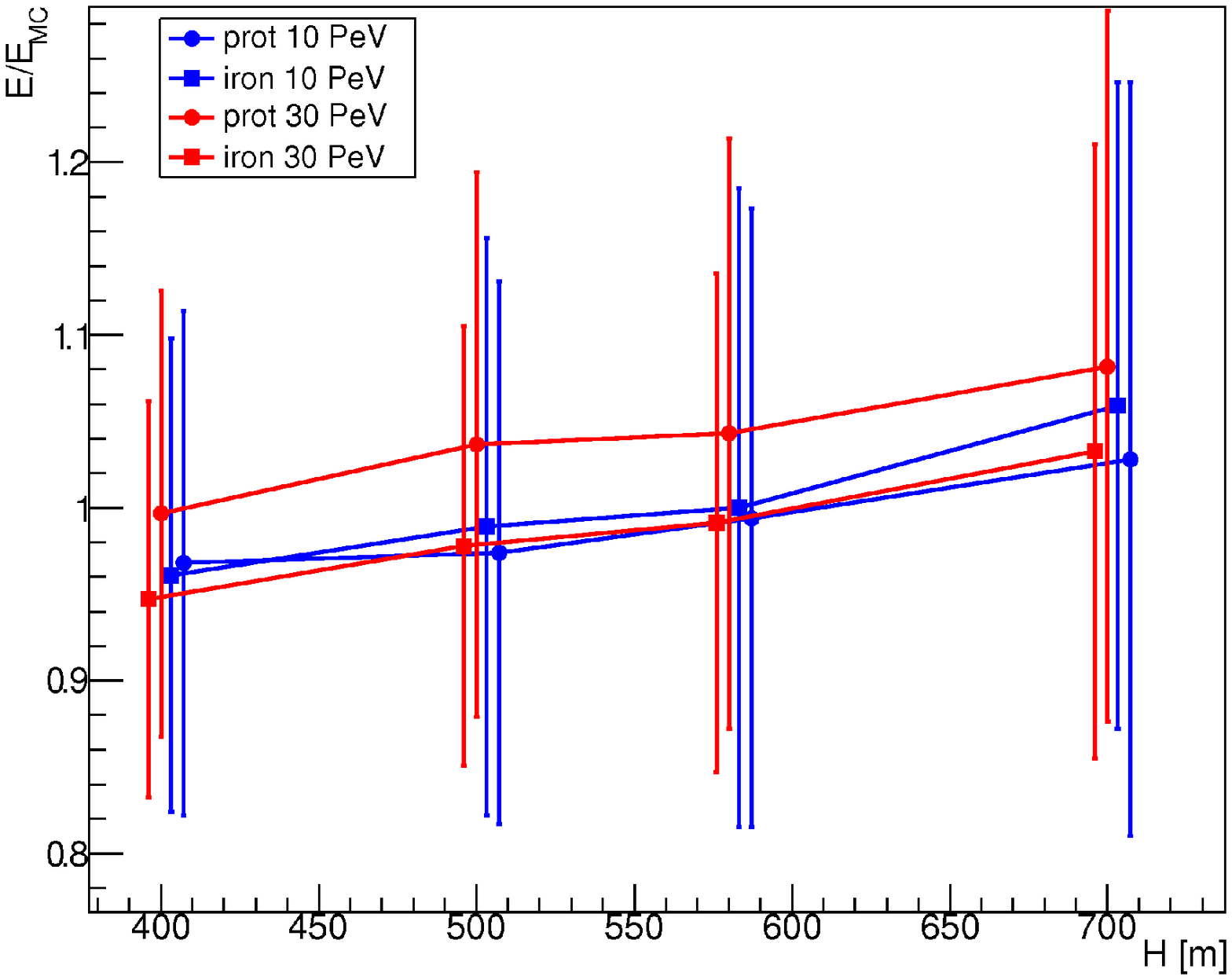}
\caption{\label{label} The same, as in figure 2, but only for showers that were registered by the model of the instrumental trigger.}
\end{minipage} 
\end{figure}

\begin{figure}[!p]
\begin{minipage}{17pc}
\includegraphics[width=17pc]{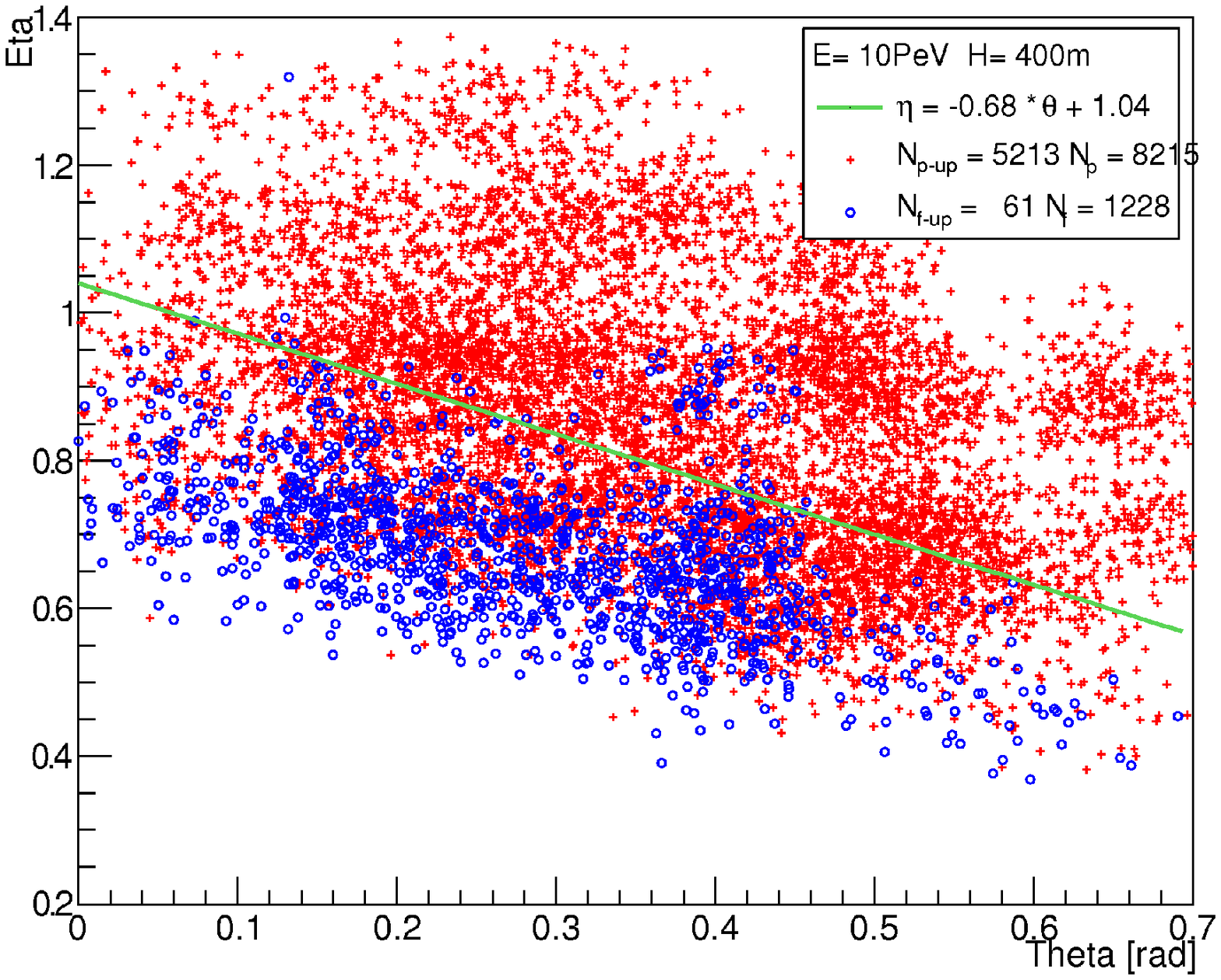}
\caption{\label{label} The result of the model showers classification for $E_{Trig}$= 11.2 PeV.}
\end{minipage}\hspace{4pc}%
\begin{minipage}{17pc}
\includegraphics[width=17pc]{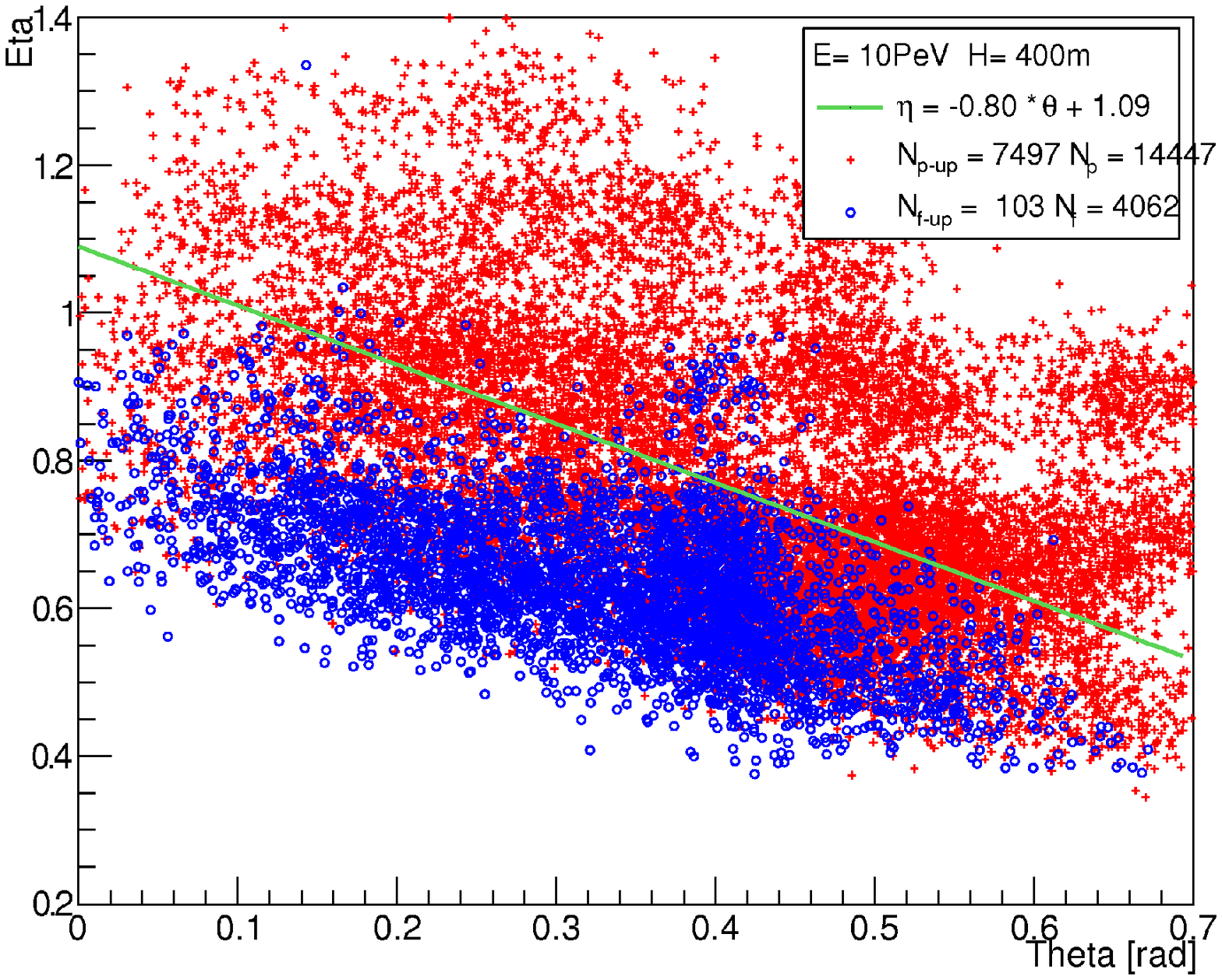}
\caption{\label{label} The same, as in figure 4, but for $E_{Trig}$= 15.9 PeV.}
\end{minipage} 
\end{figure}

\begin{figure}[!p]
\begin{minipage}{17pc}
\includegraphics[width=17pc]{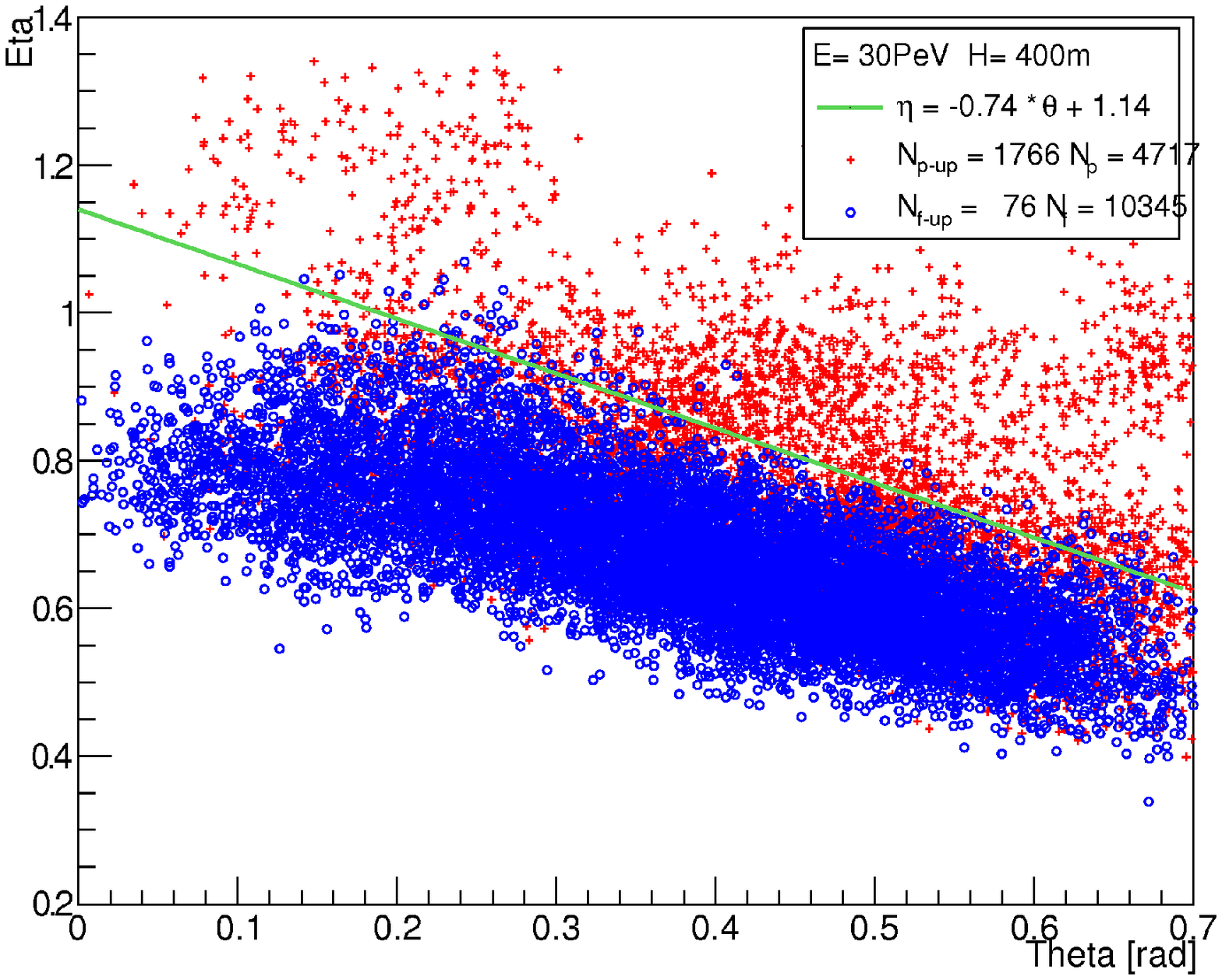}
\caption{\label{label} The same, as in figures 4-5, but for $E_{Trig}$= 20.9 PeV.}
\end{minipage}\hspace{4pc}%
\begin{minipage}{17pc}
\includegraphics[width=17pc]{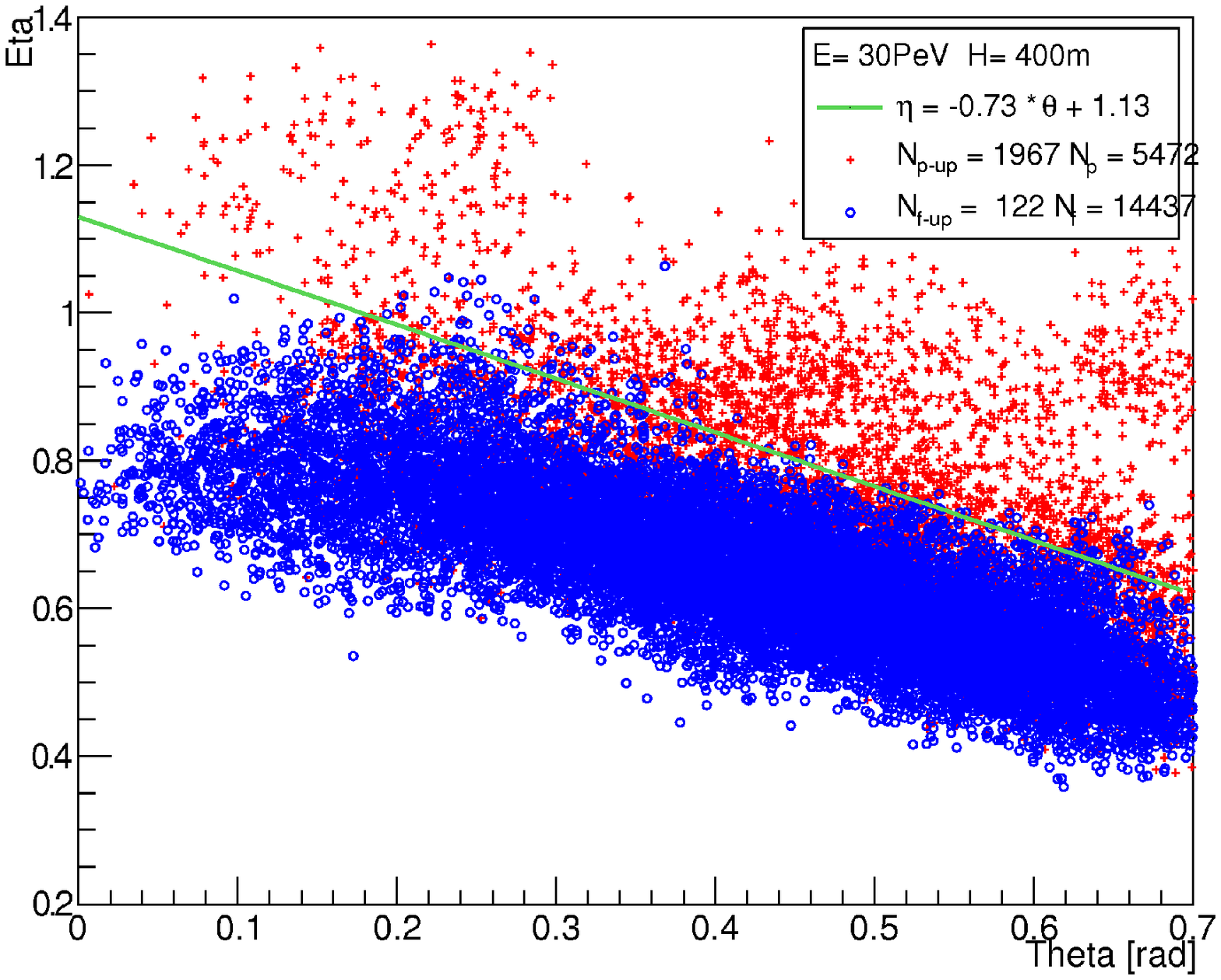}
\caption{\label{label} The same, as in figures 4-6, but for $E_{Trig}$= 29.6 PeV.}
\end{minipage} 
\end{figure}

\begin{figure}[!p]
\begin{minipage}{17pc}
\includegraphics[width=17pc]{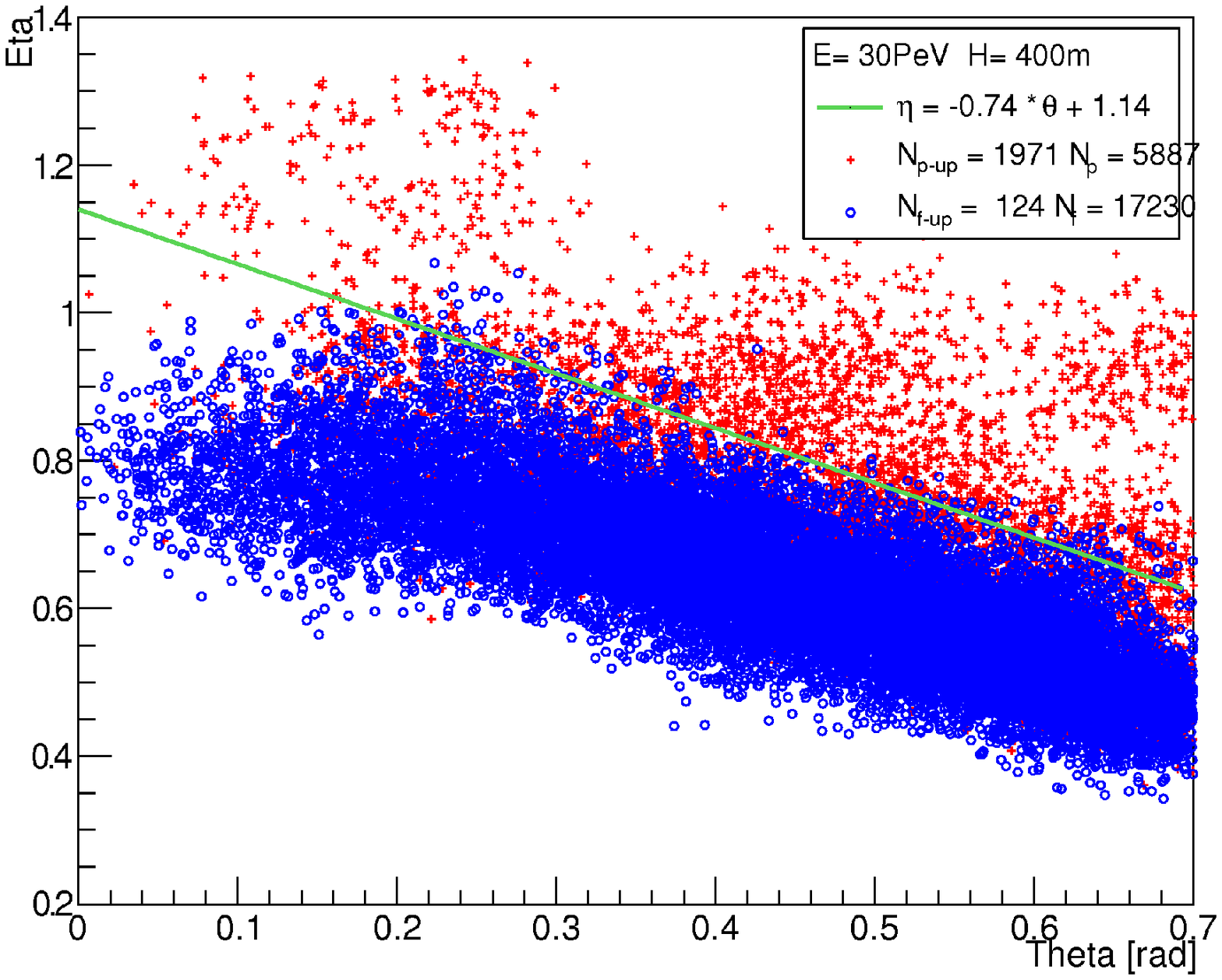}
\caption{\label{label} The same, as in figures 4-7, but for $E_{Trig}$= 41.8 PeV.}
\end{minipage}\hspace{4pc}%
\begin{minipage}{17pc}
\includegraphics[width=17pc]{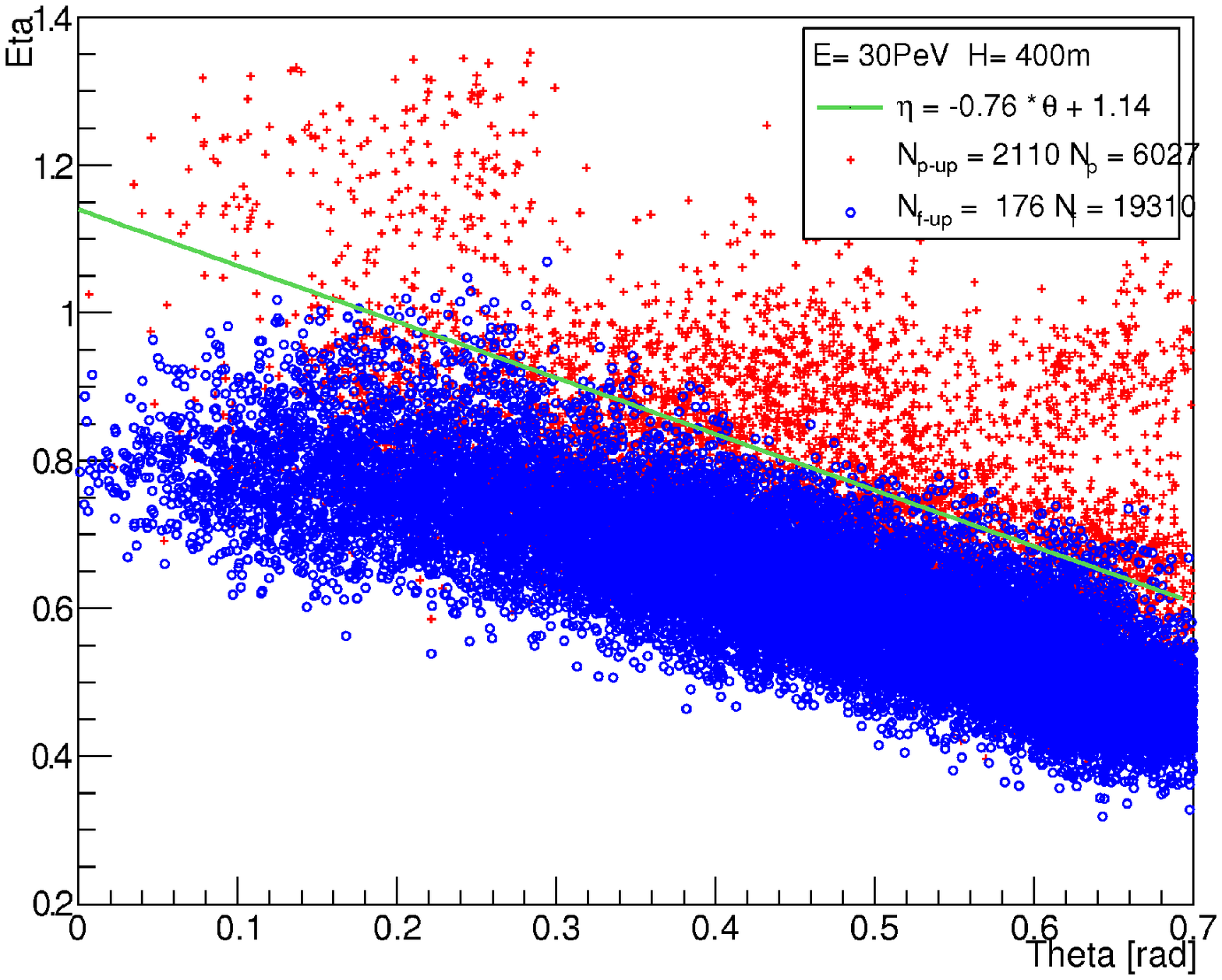}
\caption{\label{label} The same, as in figures 4-8, but for $E_{Trig}$= 102.6 PeV.}
\end{minipage} 
\end{figure}

As a first step of the elemental groups separation procedure, we have estimated the LDF steepness parameter $\eta$ for all model showers, integrated on time, by fitting these "discrete LDF" by a set of specially calculated smooth continious curves. Each curve of the set (we call it a "composite model LDF", CLDF) was composed from many model "discrete LDF" corresponding to the same CORSIKA shower. An example of a CLDF is shown in figure 1 by green circles, and the corresponding individual model LDF is plotted by red circles. The normalisation factor of a CLDF, that is nearly proportional to the primary energy \cite{ded04}, was utilized as the primary energy estimator, and the CLDF steepness parameter --- as an estimate of the $\eta$ value. The regions (0, 70) $m$ and (70, 140) $m$ used for the $\eta$ parameter estimation are constrained by vertical black lines in figure 1.

The energy reconstruction performances of the described procedure are presented in figures 2-3. Figure 2 deals with all showers in the model sample, irrespectively of the axis position. The mean reconstructed energy values are shown by symbols for proton and Iron primaries for the primary ("true MC") energy values $E_{MC}$= (10, 30, 100) $PeV$, and several altitudes (400, 500, 580, 700, 900) $m$. The corresponding standard deviations $\sigma_{E}$ ($\sigma_{E}$ is a measure of the primary energy statistical reconstruction uncertainty) are shown by bars. A small artificial shift between the results for different conditions is introduced on the $H$ axis to make them visible. Similar results for $E_{MC}$= 10 $PeV$ and 30 $PeV$ are shown in figure 3 for showers that satisfy the trigger condition. We do not show results for 900 $m$, because the maximal altitude for the 2013 run was $\approx$700 $m$. It is evident that the statistical uncertainty of the primary energy value is qiute low (from $\sim$10 \% to $\sim$20 \% depending from the primary energy and the altitude values), and the systematic error is moderate ($<$5 \% for most of the cases).

\subsection{Separation of the elemental groups}

To study the CR composition, we employed a set of Bayesian classifiers \cite{the03} using the above-described model LDF sample as a training set. In the present work, for simplicity, we constrain ourselves to the case of the linear border between the classes of primary nuclei, and two event classes --- proton and Iron primary nuclei. Following the famous "Bayes rule", we have set the prior probabilities 0.5 for both classes. The results of the classifier training are shown for the altitude value $H$= 400 $m$ and different energies $E_{Trig}$ in figures 4-9. The values of the LDF steepness parameter for proton (red crosses) and Iron (blue circles) vs. the primary zenith angle value $\theta$ are plotted; the boundary between the classes is shown by green straight line. The $E_{MC}$ parameter value is shown (see subsection 3.1 for explanation), as well as the parameters of the border line equation. The numbers of proton ($N_{p-up}$ and $N_{p}$) and Iron ($N_{f-up}$ and $N_{f}$) showers above the border, and total numbers of these model showers, respectively, are presented. To suppress statistical fluctuations arising from a small number of CORSIKA showers, we have introduced a small Gaussian smearing on both parameters $(\theta,\eta)$.

The fraction of model proton showers classified as light nuclei is $F_{p}= N_{p-up}/N_{p}$= (0.635, 0.519, 0.374, 0.359, 0.335, 0.350) for the $E_{Trig}$ values (11.2, 15.9, 20.9, 29.6, 41.8, 102.6) $PeV$, respectively, and the corresponding fraction of Iron contamination $F_{f}= N_{f-up}/N_{f}$= (0.050, 0.025, 0.007, 0.008, 0.007, 0.009). It is evident from figures 4-5 that Iron nuclei are effectively suppressed by the instrumental trigger effects; at sufficiently low energy the SPHERE-2 detector could observe practically only light nuclei. Overall, the Iron contamination appears to be quite low. Results such as shown in figures 4-9 were obtained for 5 altitudes (400, 500, 580, 700, 900) $m$ and 54 logarithmically spaced energy bins from 5 to 200 $PeV$. 

All experimental showers from the sample defined in section 2 have the parameters $(\theta,E,\eta)$ estimated, and the altitude value $H$ measured. These events were classified as proton-like or Iron-like according to their position above or below the border. The total number of the light CR component showers in 6 energy bins in the energy range 5-200 $PeV$ was reconstructed as $N_{i-light}= N_{i-p-like}/F_{p}$, where $N_{i-p-like}$ is the number of experimental showers above the border in the $i$th bin. The light and heavy component intensities were corrected for the instrumental acceptance effects to account for the difference between the detection efficiencies of light and heavy nuclei.

In comparison to \cite{ant13}, in the present work we were able to substantially lower the threshold of the method using calculations of the instrumental acceptance \cite{ant13b},\cite{ant15a}. The first-order correction in the present work uses the 2013-integrated acceptance curves for the case of $R< R_{FOV}+100$ $m$, instead of $R< R_{FOV}+30$ $m$. The number of available model showers at $E$= 100 $PeV$ is currently low, therefore the analysis was performed with the 10 $PeV$ and 30 $PeV$ model showers. Corrections for these inaccuracies are made in section 4.

\section{Results}

\begin{figure}[h]
\begin{minipage}{18pc}
\includegraphics[width=18pc]{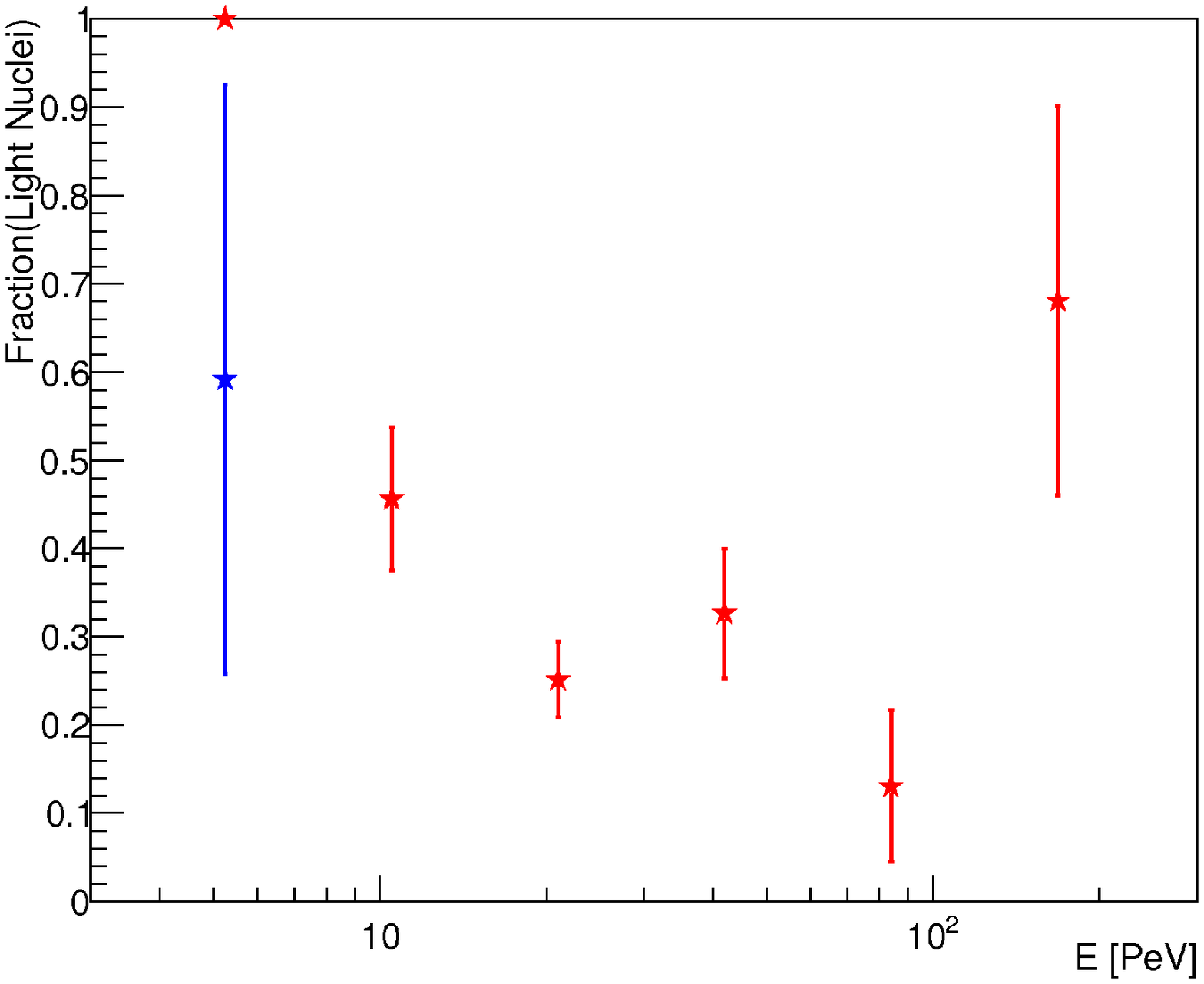}
\caption{\label{label} The fraction of light ("proton-like") nuclei reconstructed using the 2013 run data of the SPHERE experiment.}
\end{minipage}\hspace{2pc}%
\begin{minipage}{18pc}
\includegraphics[width=18pc]{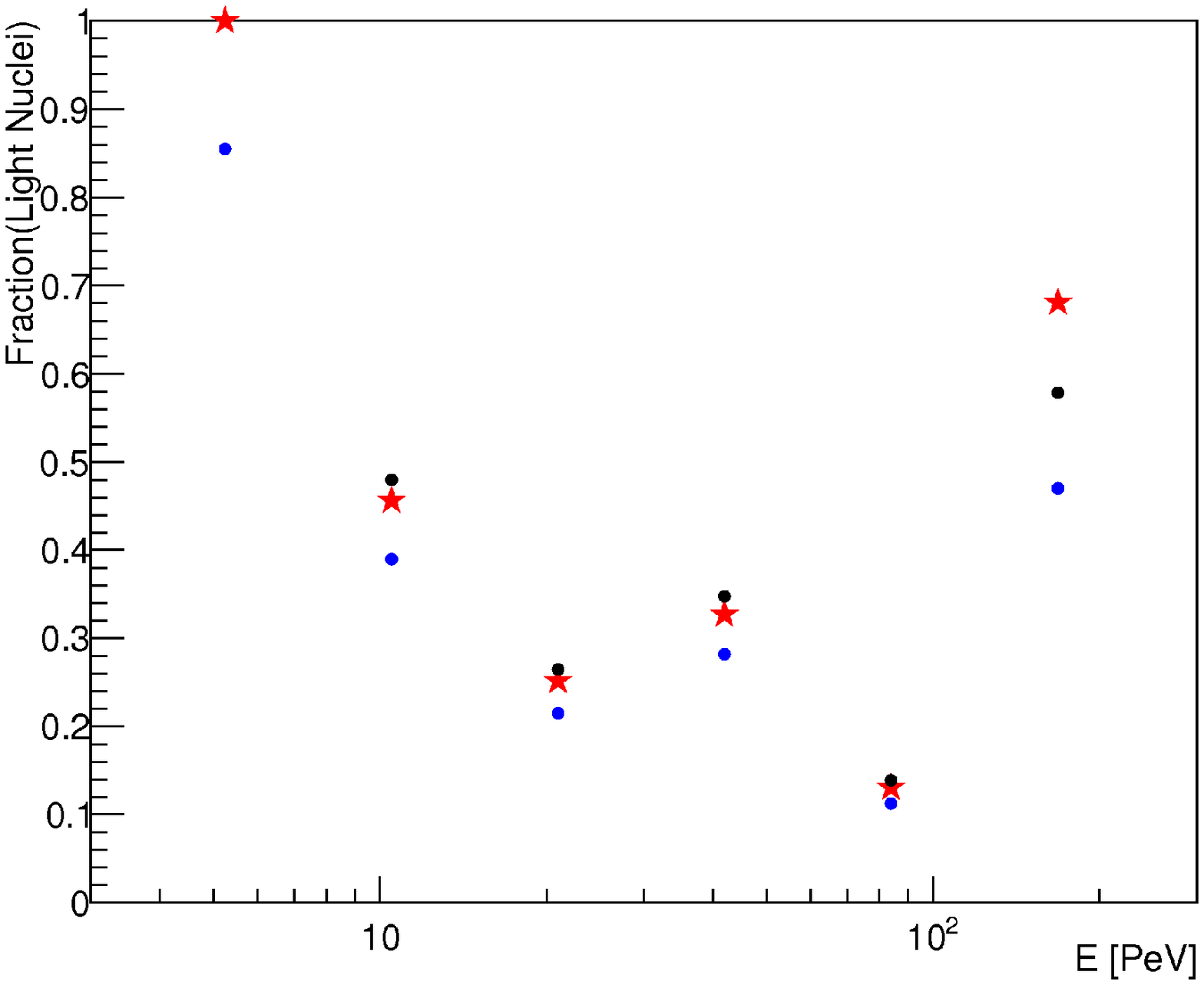}
\caption{\label{label} The fraction of light nuclei before (red stars) and after corrections (black and blue circles; see text for more details).}
\end{minipage} 
\end{figure}

The result of our analysis --- the estimated fraction of light nuclei in the energy range 5-200 $PeV$ --- is shown in figure 10 by red stars with statistical uncertainties. For the case of the 120--240 $PeV$ energy interval the statistical uncertainty is quite high and our results are not constraining. For the case of 3.5--7 $PeV$ energy interval we didn't find any indication of the Iron nuclei presence. To estimate the statistical uncertainty for this bin, we have added one "bonus" Iron shower; the corresponding result is shown in figure 10 by the blue star. The light nuclei fraction behaviour presented in figure 10 is qualitatively similar to the KASCADE-Grande result \cite{ape13b}, and doesn't contradict our earlier estimate \cite{ant13} obtained with the 2012 run data. We have already applied the correction for model energy discreteness here (see section 3).

Several systematic effects do exist that could change our results. A shift of the light nuclei fraction could occur as a result of the energy estimation systematic uncertainty (see figure 3). For some energy bins, this effect tends to enhance the reconstructed light nuclei intensity up to 10--15 \% with respect to its true value. Therefore, the corrected fraction of light nuclei in these bins would be lower than plotted in figure 10 (for instance, 25-27 \% instead of 30 \%). We have estimated and applied corrections for this effect, as well as the acceptance correction discussed at the end of section 3. The new result after these corrections is shown in figure 11 by black circles. As well, we have estimated the LDF reconstruction procedure systematic uncertainty, and the corresponding result including this last correction, as well as all previous ones, is shown in figure 11 by blue circles. Dedicated work to study these and other systematic effects is in progress, and will be reported elsewhere.

\section{Multidimensional method for superhigh energy CR composition study}

In sections 3--4 were have employed the set of one-dimensional criteria, relying on the $\eta= \eta(A,E_{Trig},\theta,H)$ parameter. Here we show that it is possible to improve the nuclei classes separability by using multidimensional pattern recognition methods. We demonstrate the case on our model sample (without imposing the trigger effects) by studiyng classification performances for 3 nuclei groups separation --- proton, Nitrogen, and Iron. The feature vector was chosen to be 4-dimensional, and was composed of Cherenkov light intensities in concentric rings with the center in the shower's axis and with radii (0, 40) $m$, (40, 80) $m$, (80, 120) $m$, and (120, 160) $m$. The zenith angle value $\theta$ was added as the fifth classification parameter. The multivariate Bayesian pattern recognition technique under assumption of multidimensional Gaussian distribution of features \cite{the03} was used for this study. This technique was recently shown to be very effective for the case of CR composition study with lateral-angular distribution of Cherenkov light \cite{gal11}--\cite{bor13}.

\subsection{Proton--Nitrogen separation}

All graphs in this subsection (figures 12-15) show contamination of Nitrogen nuclei (i.e. the fraction of Nitrogen nuclei classified as protons) vs. proton selection efficiency. Black curves are drawn for $H$= 400 $m$, red curves --- for $H$= 580 $m$, and blue curves --- for $H$= 900 $m$. The primary energy values and zenith angle range are shown in the captions to the graphs.

\begin{figure}[h]
\begin{minipage}{18pc}
\includegraphics[width=18pc]{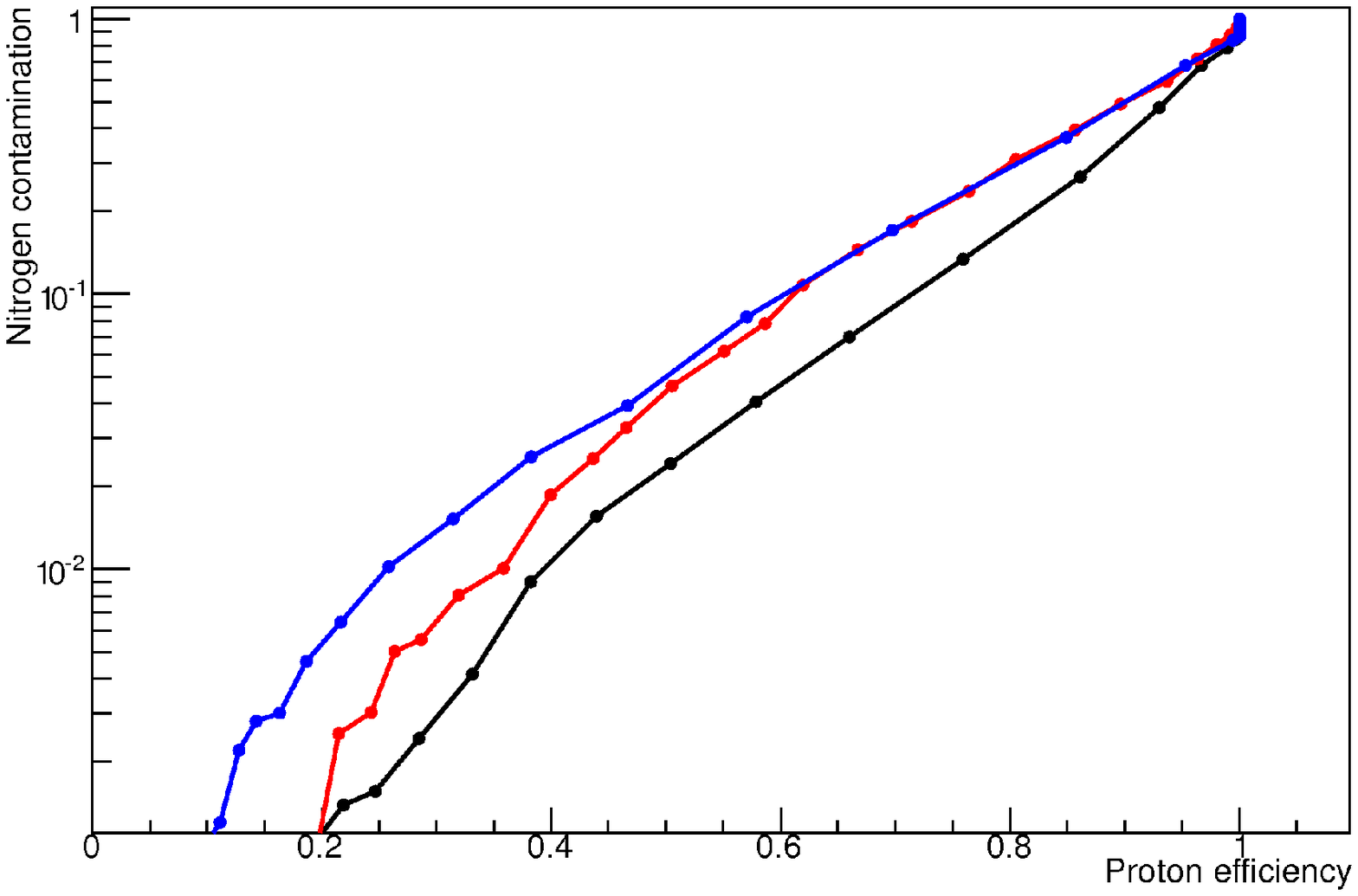}
\caption{\label{label} Proton-Nitrogen separation for $E$= 10 $PeV$ and $\theta$= 0-20$^{\circ}$.}
\end{minipage}\hspace{2pc}%
\begin{minipage}{18pc}
\includegraphics[width=18pc]{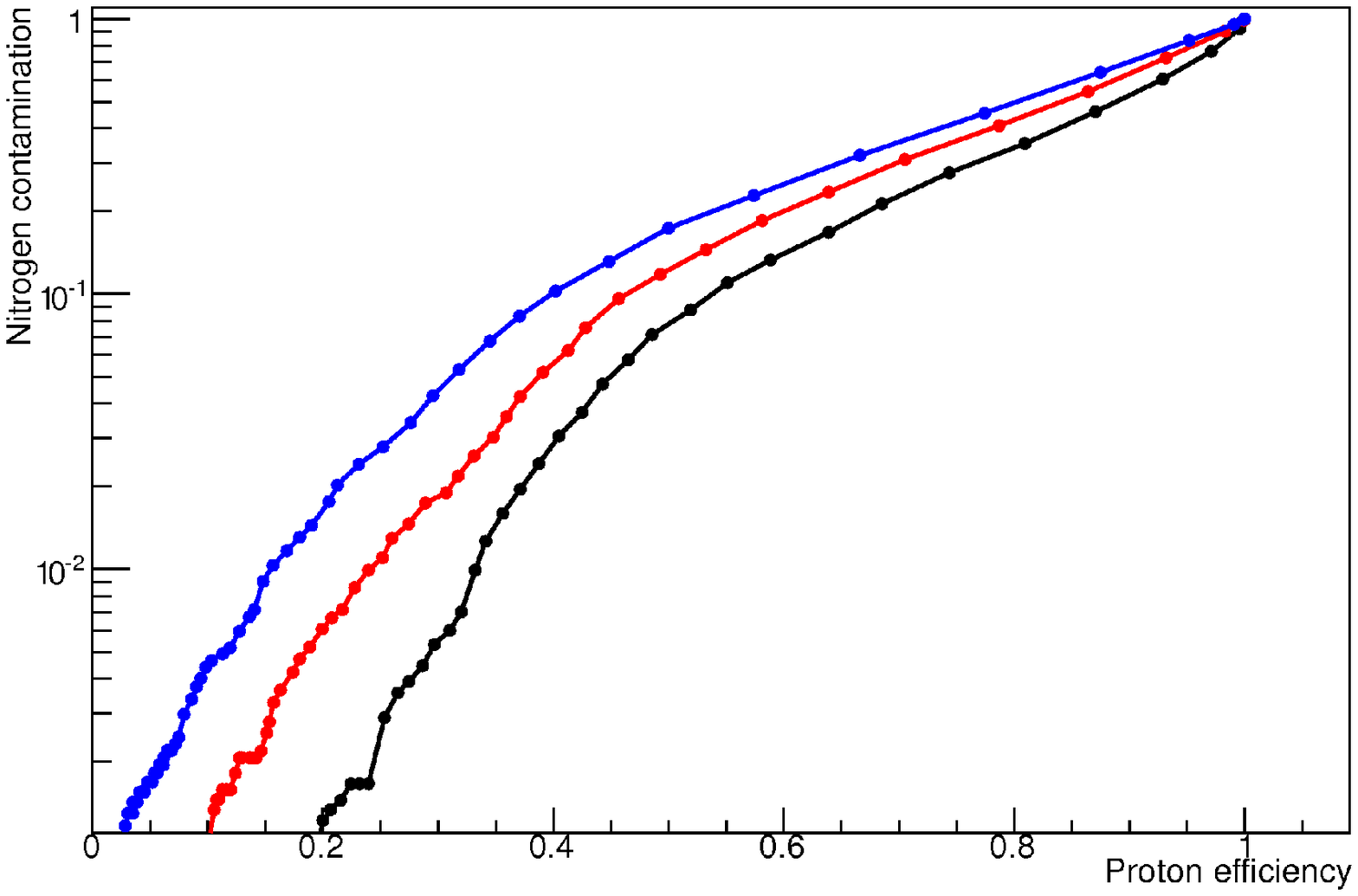}
\caption{\label{label} The same, as in figure 12, but for $\theta$= 20-40$^{\circ}$.}
\end{minipage} 
\end{figure}

\begin{figure}[h]
\begin{minipage}{18pc}
\includegraphics[width=18pc]{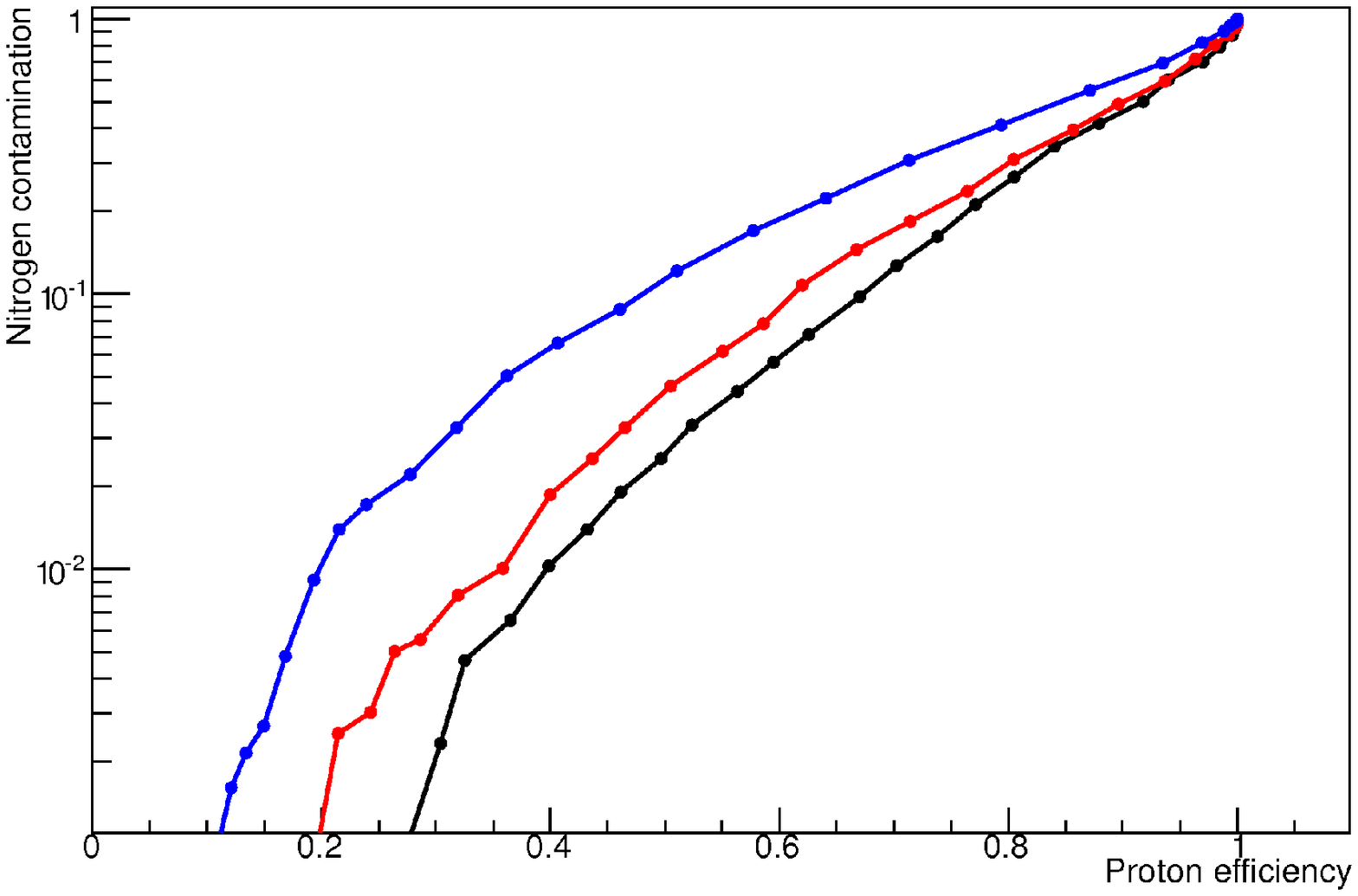}
\caption{\label{label} The same, as in figures 12-13, but for $E$= 30 $PeV$ and $\theta$= 0-20$^{\circ}$.}
\end{minipage}\hspace{2pc}%
\begin{minipage}{18pc}
\includegraphics[width=18pc]{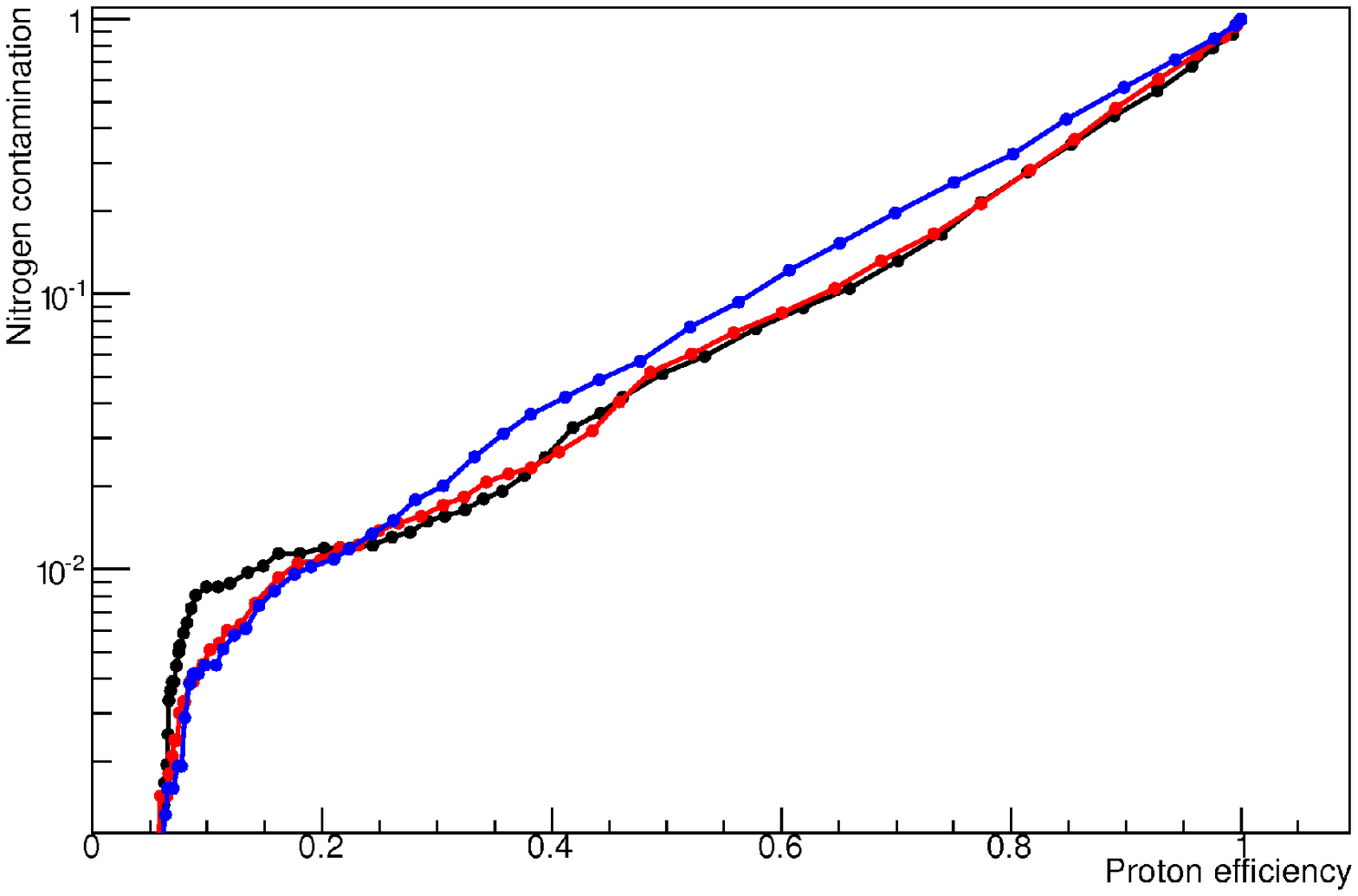}
\caption{\label{label} The same, as in figure 14, but for $\theta$= 20-40$^{\circ}$.}
\end{minipage} 
\end{figure}

\subsection{Proton--Iron, proton--Helium and Nitrogen--Iron separation}

The same results as in figures 12-15, but for separation of other elemental groups, were also obtained. For the case of proton-Iron separation (except the 10 $PeV$ and $\theta$= 20-40$^{\circ}$ case), the typical proton selection efficiency corresponding to $\sim$1 \% Iron contamination is $\approx$60-90 \%, depending on conditions. For the 10 $PeV$ and $\theta$= 20-40$^{\circ}$ case, the proton selection efficiency is lower, but in this case the Iron showers would be suppressed by the trigger effects, as was explained in section 3. The same applies to other cases where proton selection efficiency is $<$75 \%; therefore, the proton selection efficiency is, in fact, $>$75 \% for $\sim$1 \% Iron contamination.

As well, we have studied the Nitrogen--Iron classification performance, and obtained the typical Nitrogen selection efficiency value $\approx$40 \% for $\sim$1 \% Iron contamination. To conclude, let us note that proton--Helium separation is, in principle, also possible with the presented multidimensional technique. However, with the criteria considered in the present work, the proton selection efficiency was observed to be always $<$25 \% for $\sim$1 \% Helium contamination.

\section{Conclusions}

We have presented and discussed the method for event-by-event study of superhigh CR composition applicable to the SPHERE experiment working conditions. It was found that a part of the light component showers can be readily selected against the heavy component background. The primary energy can be reconstructed with sufficiently low uncertainty, both statistical and systematic. The result of the analysis --- the light nuclei fraction vs. energy --- is qualitatively similar to the KASCADE-Grande result, and doesn't contradict to our earlier estimates. As well, we have shown that multidimensional criteria, in principle, allow to enhance the separability of the elemental group classes.

\subsection*{Acknowledgments}
The authors are grateful to the technical collaborators of the SPHERE-2 experiment. The work was supported by the Russian Foundation for Basic Research (grants 11-02-01475-a, 12-02-10015-k, 13-02-00470-а); the Russian President grants LSS-871.2012.2; LSS-3110.2014.2; the Program of basic researches of Presidium of Russian Academy of Sciences "Fundamental properties of a matter and astrophysics". Calculations were performed using the SINP MSU space monitoring data center computer cluster; we are grateful to Dr. V.V. Kalegaev for permission to use the hardware and to V.O. Barinova, M.D. Nguen, D.A. Parunakyan for technical support. The reported study was supported by the Supercomputing Center of Lomonosov Moscow State University.


\section*{References}

\begin{thebibliography}{999}
\bibitem{tsu08}
 Tsunesada Y et al. (BASJE) 2008 {\it Proc. Int. Conf. on Cosmic Rays} vol 4 127
\bibitem{kam12}
 Kampert K-H and Unger M 2012 {\it APh} {\bf 35} 660
\bibitem{ant05}
Antoni T et al. (KASCADE) 2005 {\it APh} {\bf 24} 1
\bibitem{ant09}
Antoni T et al. (KASCADE) 2009 {\it APh} {\bf 31} 86
\bibitem{gar07}
Garyaka A et al. (GAMMA) 2007 {\it APh} {\bf 28} 169
\bibitem{ape13a}
Apel W D et al. (KASCADE-Grande) 2013 {\it APh} {\bf 47} 54
\bibitem{ape13b}
Apel W D et al. (KASCADE-Grande) 2013 {\it Phys. Rev.} {\bf D87} 081101(R)
\bibitem{ape14}
Apel W D et al. (KASCADE-Grande) 2014 {\it Advances in Space Research} {\bf 53} 1456
\bibitem{ant02}
Antoni T et al. (KASCADE-Grande) 2002 {\it APh} {\bf 16} 245
\bibitem{ter07}
Ter-Antonyan S V 2007 {\it APh} {\bf 28} 321
\bibitem{ant15a}
Antonov R A and et al. 2015 {\it Physics of Particles and Nuclei} {\bf 46} 60
\bibitem{chu74}
Chudakov A E 1974 {\it Proc. All-USSR Symp. on Exp. Meth. of UHECR, Yakutsk} (Yakutsk: Russian Academy of Sciences Publisher) (In Russian) 69 
\bibitem{ant15b}
Antonov R A and et al. 2015  {\it  DESY Conf. Proc. (PANIC-2014)} (In print)
\bibitem{ant13}
Antonov R A and et al. 2013 {\it J. Phys.: Conf. Series} {\bf 409} 012088 
\bibitem{ano09}
Anokhina A M and et al. 2009 {\it Bulletin of the Lebedev Physics Institute} {\bf 36} 146
\bibitem{ant09}
Antonov R A and et al. 2009 {\it Proc. Int. Conf. on Cosmic Rays (Lodz)} vol HE.1.3 (Lodz:) id. 434
\bibitem{hec98}
Heck D and et al. 1998 {\it FZKA Report} {\bf 6019}
\bibitem{kal97}
Kalmykov N et al. 1997 {\it Nucl. Phys. Proc. Suppl.} B {\bf 52} 17
\bibitem{fes85}
Fesefeldt H C 1985 {\it Technical Report} {\bf PITHA 85-02 RWTH}
\bibitem{ago03}
Agostinelli S and et al. 2003 {\it NIM} A {\bf 506} 250
\bibitem{ded04}
Dedenko L G and et al. 2004 {\it Nucl. Phys.: Proc. Suppl.} B {\bf 136} 1217
\bibitem{the03}
Theodoridis S and Koutroumbas K 2003 {\it Pattern Recognition} 2nd edition (Amsterdam: Elsevier)
\bibitem{ant13b}
Antonov R A and et al. 2013 {\it Proc. Int. Conf. on Cosmic Rays (Rio de Janeiro)} (Rio de Janeiro:) id. 1185
\bibitem{gal11}
Galkin V I and Dzhatdoev T A 2011 {\it Izvestiya Rossiiskoi Akademii Nauk} {\bf 75} 338 (In Russian)
\bibitem{bor13}
Borisov A S and  Galkin V I 2013 {\it J. Phys.: Conf. Series} {\bf 409} 012089
\end{thebibliography}
\end{document}